\documentclass[aps,prl,preprint,superscriptaddress,showpacs,byrevtex]{revtex4}
\usepackage{graphicx}
\usepackage{epsfig}
\usepackage{dcolumn}
\usepackage{bm,color}

\newcommand{\MMS}{M_{\rm rec}^2}
\newcommand{\y}{Y(4260)}
\newcommand{\lum}{{\cal L}}
\newcommand{\eff}{\varepsilon}
\newcommand{\BR}{{\cal B}}

\newcommand{\pip}{\pi^+}
\newcommand{\pim}{\pi^-}

\newcommand{\kap}{K^+}
\newcommand{\kam}{K^-}
\newcommand{\ks}{K_S^0}

\newcommand{\psp}{\psi(2S)}

\newcommand{\jpsi}{J/\psi}

\newcommand{\psiftf}{\psi(4415)}
\newcommand{\EE}{e^+e^-}
\newcommand{\MM}{\mu^+\mu^-}
\newcommand{\LL}{\ell^+\ell^-}
\newcommand{\pp}{\pi^+\pi^-}
\newcommand{\kk}{K^+K^-}
\newcommand{\ksks}{K_S^0K_S^0}

\newcommand{\ppjpsi}{\pi^+\pi^- J/\psi}
\newcommand{\pppsp}{\pi^+\pi^- \psp}
\newcommand{\kkjpsi}{K^+K^- J/\psi}
\newcommand{\ksksjpsi}{K_S^0K_S^0 J/\psi}

\newcommand{\beq}{\begin{equation}}
\newcommand{\eeq}{\end{equation}}
\newcommand{\bitm}{\begin{itemize}}
\newcommand{\eitm}{\end{itemize}}

\def\Journal#1#2#3#4{{#1} {\bf #2}, #3 (#4)}

\def\PRL{Phys. Rev. Lett.}
\def\PRD{Phys. Rev. D}

\begin{document}


\title{
Observation of $\EE \to \kkjpsi$ via Initial State Radiation at
Belle}

\affiliation{Budker Institute of Nuclear Physics, Novosibirsk}
\affiliation{Chiba University, Chiba} \affiliation{University of
Cincinnati, Cincinnati, Ohio 45221}
\affiliation{The Graduate University for Advanced Studies, Hayama}
\affiliation{Hanyang University, Seoul} \affiliation{University of
Hawaii, Honolulu, Hawaii 96822} \affiliation{High Energy
Accelerator Research Organization (KEK), Tsukuba}
\affiliation{Hiroshima Institute of Technology, Hiroshima}
\affiliation{Institute of High Energy Physics, Chinese Academy of
Sciences, Beijing} \affiliation{Institute of High Energy Physics,
Vienna} \affiliation{Institute of High Energy Physics, Protvino}
\affiliation{Institute for Theoretical and Experimental Physics,
Moscow} \affiliation{J. Stefan Institute, Ljubljana}
\affiliation{Kanagawa University, Yokohama} \affiliation{Korea
University, Seoul}
\affiliation{Kyungpook National University, Taegu}
\affiliation{\'Ecole Polytechnique F\'ed\'erale de Lausanne
(EPFL), Lausanne} \affiliation{University of Ljubljana, Ljubljana}
\affiliation{University of Maribor, Maribor}
\affiliation{University of Melbourne, School of Physics, Victoria
3010} \affiliation{Nagoya University, Nagoya} \affiliation{Nara
Women's University, Nara} \affiliation{National Central
University, Chung-li} \affiliation{National United University,
Miao Li} \affiliation{Department of Physics, National Taiwan
University, Taipei} \affiliation{H. Niewodniczanski Institute of
Nuclear Physics, Krakow} \affiliation{Nippon Dental University,
Niigata} \affiliation{Niigata University, Niigata}
\affiliation{University of Nova Gorica, Nova Gorica}
\affiliation{Osaka City University, Osaka} \affiliation{Osaka
University, Osaka} \affiliation{Panjab University, Chandigarh}
\affiliation{Saga University, Saga} \affiliation{University of
Science and Technology of China, Hefei} \affiliation{Seoul
National University, Seoul}
\affiliation{Sungkyunkwan University, Suwon}
\affiliation{University of Sydney, Sydney, New South Wales}
\affiliation{Tata Institute of Fundamental Research, Mumbai}
\affiliation{Toho University, Funabashi} \affiliation{Tohoku
Gakuin University, Tagajo} \affiliation{Tohoku University, Sendai}
\affiliation{Department of Physics, University of Tokyo, Tokyo}
\affiliation{Tokyo Institute of Technology, Tokyo}
\affiliation{Tokyo Metropolitan University, Tokyo}
\affiliation{Tokyo University of Agriculture and Technology,
Tokyo}
\affiliation{Virginia Polytechnic Institute and State University,
Blacksburg, Virginia 24061}
\affiliation{Yonsei University, Seoul}
  \author{C.~Z.~Yuan}\affiliation{Institute of High Energy Physics, Chinese Academy of Sciences, Beijing} 
  \author{C.~P.~Shen}\affiliation{Institute of High Energy Physics, Chinese Academy of Sciences, Beijing} 
  \author{P.~Wang}\affiliation{Institute of High Energy Physics, Chinese Academy of Sciences, Beijing} 
  \author{X.~L.~Wang}\affiliation{Institute of High Energy Physics, Chinese Academy of Sciences, Beijing} 
  \author{I.~Adachi}\affiliation{High Energy Accelerator Research Organization (KEK), Tsukuba} 
  \author{H.~Aihara}\affiliation{Department of Physics, University of Tokyo, Tokyo} 
  \author{K.~Arinstein}\affiliation{Budker Institute of Nuclear Physics, Novosibirsk} 
  \author{V.~Aulchenko}\affiliation{Budker Institute of Nuclear Physics, Novosibirsk} 
  \author{T.~Aushev}\affiliation{\'Ecole Polytechnique F\'ed\'erale de Lausanne (EPFL), Lausanne}\affiliation{Institute for Theoretical and Experimental Physics, Moscow} 
  \author{A.~M.~Bakich}\affiliation{University of Sydney, Sydney, New South Wales} 
% ...
\collaboration{The Belle Collaboration}

\date{\today}

\begin{abstract}

The process $\EE\to \kkjpsi$ is observed for the first time via
initial state radiation. The cross section of $\EE\to \kkjpsi$ for
center-of-mass energies between threshold and 6.0~GeV is measured
using 673~fb$^{-1}$ of data collected with the Belle detector on
and off the $\Upsilon(4S)$ resonance. No significant signal for
$\y\to \kk\jpsi$ is observed, and we determine $\BR(\y\to
\kkjpsi)\Gamma(\y\to\EE)<1.2$~eV/$c^2$ at a 90\% confidence level.
We also find evidence for $\EE\to \ksksjpsi$ in the same data
sample.

\end{abstract}


\maketitle

The study of charmonium states via initial state radiation ($ISR$)
at the $B$-factories has proven to be very fruitful. In the
process $\EE \to \gamma_{ISR} \ppjpsi$, the BaBar Collaboration
observed the $\y$~\cite{babary}. This structure was also observed
by the CLEO~\cite{cleoy} and Belle Collaborations~\cite{belley}
with the same technique; moreover, there is a broad structure near
4.05~GeV/$c^2$ in the Belle data. In a subsequent search for the
$\y$ in the $\EE \to \gamma_{ISR} \pppsp$ process, BaBar found a
structure at around 4.32~GeV/$c^2$~\cite{babar_pppsp}, while the
Belle Collaboration observed two resonant structures at
4.36~GeV/$c^2$ and 4.66~GeV/$c^2$~\cite{belle_pppsp}. Recently,
CLEO collected 13.2~pb$^{-1}$ of data at $\sqrt{s}=4.26$~GeV and
investigated 16 decay modes with charmonium or light
hadrons~\cite{cleoy4260}. The large $\EE\to \ppjpsi$ cross section
at this energy is confirmed. In addition, there is also evidence
for $\kk \jpsi$ (3.7$\sigma$) based on three events observed.
Further investigation on the process $\EE\to \kkjpsi$ will shed
light on the understanding of the $\y$ and the other vector
charmonium states.

In this Letter, we use a 673~fb$^{-1}$ data sample collected near
the $\Upsilon(4S)$ with the Belle detector~\cite{Belle} operating
at the KEKB asymmetric-energy $e^+e^-$ (3.5 on 8~GeV)
collider~\cite{KEKB} to investigate the $\kkjpsi$ final state
produced via $ISR$. About 90\% of the data were collected at the
$\Upsilon(4S)$ resonance ($\sqrt{s}=10.58$~GeV), and the rest were
taken at a center-of-mass (CM) energy that is 60~MeV below the
$\Upsilon(4S)$ peak.

We use PHOKHARA~\cite{phokhara} that was validated in previous
analysis~\cite{belley} to generate signal events. In the
generator, one or two photons are allowed to be emitted before
forming the resonance $X$, then $X$ decays into $\kkjpsi$ with
$\jpsi$ decays into $\EE$ or $\MM$. When generating the MC sample,
the mass of the $X$ is fixed to a certain value while the width is
set to zero. In $X\to \kkjpsi$, a pure $S$-wave between the $\kk$
system and the $\jpsi$, as well as between the $\kap$ and $\kam$
is assumed. The invariant mass of the $\kk$ system is generated
according to phase space. To estimate the model uncertainty, we
also generate events with $\kk$ invariant mass distributed like
$m_{\pip\pim}$ in $\psp\to \ppjpsi$ decays~\cite{besdist}, i.e.,
\( \frac{d\sigma}{dm_{\kk}} \propto \hbox{Phase Space} \times
(m_{\kk}^2-4m_K^2)^2 \).

We select candidate events with criteria similar to those used for
the analysis of $\EE\to \ppjpsi$~\cite{belley}. We require the
number of charged tracks to be four with a zero net charge. For
these tracks, the impact parameters perpendicular to and along the
beam direction with respect to the interaction point are required
to be less than $0.5$ and $4$~cm, respectively, and the transverse
momentum is restricted to be higher than 0.1~GeV/$c$. For each
charged track, information from different detector subsystems is
combined to form a likelihood for each particle species ($i$),
$\mathcal{L}_i$~\cite{pid}. Tracks with
$\mathcal{R}_K=\frac{\mathcal{L}_K}{\mathcal{L}_K+\mathcal{L}_\pi}>0.6$,
are identified as kaons with an efficiency of about 92\% for the
tracks of interest; about 4\% are misidentified $\pi$
tracks~\cite{pid}. Similar likelihood ratios are formed for
electron and muon identification. For electrons from $\jpsi\to
\EE$, one track should have $\mathcal{R}_e>0.95$ while the other
track has $\mathcal{R}_e>0.05$, this results in a very pure
$\jpsi\to \EE$ sample with an efficiency of 90\%; for muons from
$\jpsi\to \MM$, at least one track is required to have
$\mathcal{R}_\mu>0.95$; in cases where one of the tracks has no
muon identification (ID) information, the polar angles of the two
muon tracks in the $\kk\MM$ center-of-mass system are required to
satisfy $-0.7<\cos\theta_\mu<0.7$ based on a comparison between
data and MC simulation. The efficiency for $\jpsi\to \MM$ is 87\%.
Events with $\gamma$-conversions are removed by requiring
$\mathcal{R}_e < 0.75$ for the $\kap\kam$ tracks. For the
$\jpsi\to \EE$ mode, $\gamma$-conversion events are further
suppressed by requiring a $\kap\kam$ invariant mass greater than
1.05~GeV/$c^2$; this also removes events with a $\phi$ signal in
the final state. For the $\jpsi\to \MM$ mode, we require a $\kk$
invariant mass outside a $\pm 10$~MeV/$c^2$ interval around the
$\phi$ mass to remove events with a $\phi$ signal in the final
state, possibly produced via $\EE\to \gamma \gamma^* \gamma^* \to
\gamma \phi \LL$. The detection of the $ISR$ photon is not
required, instead, we identify $ISR$ events by the requirement
$|\MMS|<1.0~(\hbox{GeV}/c^2)^2$, where $\MMS$ is the square of the
mass that is recoiling from the four charged tracks. The $\MMS$
requirement is tighter than that in our previous
analyses~\cite{belley,belle_pppsp} as the $\MMS$ resolution has
improved, due to the lower momenta of the particles in $\kkjpsi$
as compared to the $\ppjpsi$ and $\pp\psp$ final states.

Clear $\jpsi$ signals are observed in both decay modes. We define
a $\jpsi$ signal region as
$3.06~\hbox{GeV}/c^2<m_{\ell^+\ell^-}<3.14~\hbox{GeV}/c^2$ (the
bremsstrahlung photons in the $\EE$ final state are included, and
the mass resolution is about 17~MeV/$c^2$), and $\jpsi$ mass
sidebands as
$2.91~\hbox{GeV}/c^2<m_{\ell^+\ell^-}<3.03~\hbox{GeV}/c^2$ or
$3.17~\hbox{GeV}/c^2<m_{\ell^+\ell^-}<3.29~\hbox{GeV}/c^2$; the
latter are three times as wide as the signal region.

\begin{figure}[htbp]
\psfig{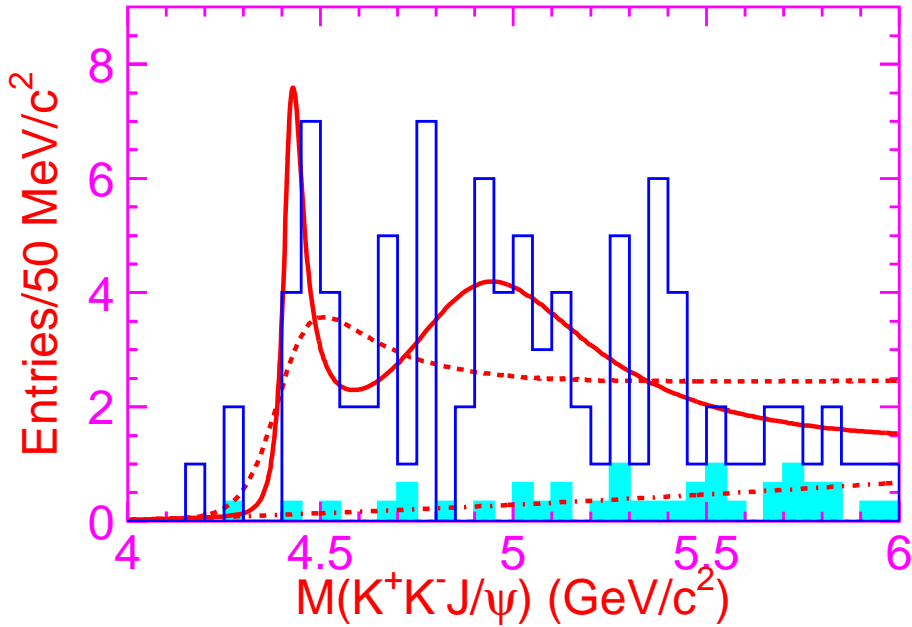}
\caption{The $\kkjpsi$ invariant mass distribution. The open
histogram is the selected data while the shaded histogram shows
the normalized sideband events. The solid (dashed) curve shows the
best fit with two (one) Breit-Wigner functions and an incoherent
background term, while the dash-dotted curve indicates a fit to
the sideband background.} \label{mkkll_full}
\end{figure}

Figure~\ref{mkkll_full} shows the $\kkjpsi$ invariant
mass~\cite{footnote} distribution after the above selection,
together with the background estimated from the $\jpsi$ mass
sidebands. There is a broad enhancement around 4.4-5.5~GeV/$c^2$.
In addition, there are two events near $\sqrt{s}=4.26$~GeV, where
CLEO observes three $\kkjpsi$ events~\cite{cleoy4260}. It is
evident from the figure that the background estimated from the
$\jpsi$ sidebands is low, which indicates that the background from
non-$\jpsi$ final states is small. The backgrounds not measured
from the sidebands, such as $X\jpsi$, with $X$ not being $\kk$,
are found from MC simulation to be less than one event and are
neglected.

\begin{figure}[htbp]
\centerline{\hbox{\psfig{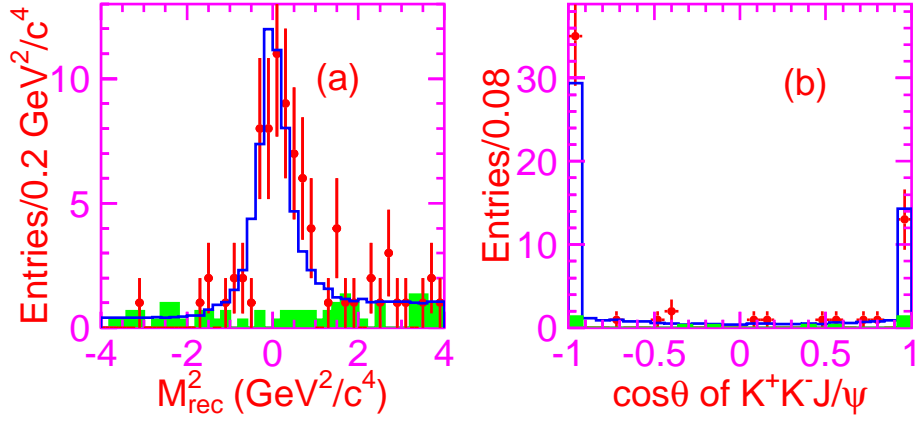}}}
\caption{$\MMS$ distribution (a) and the polar angle distribution
of the $\kkjpsi$ system in the $\EE$ CM frame (b) for the selected
$\kkjpsi$ events with invariant masses between 4.4 and
5.2~GeV/$c^2$. The points with error bars are data, the shaded
histogram is the normalized $\jpsi$ sideband distribution, and the
solid histograms are MC simulated events, normalized to the
measured cross section and integrated luminosity. The background
from $\jpsi$ mass sidebands has been added to the MC simulation.}
\label{recmx_xcosthe_y}
\end{figure}

The data points in Figs.~\ref{recmx_xcosthe_y}(a)~and~(b) show the
$\MMS$ distribution (the requirement on it has been relaxed) and
the polar angle distribution of the $\kkjpsi$ system in the $\EE$
CM frame for the selected $\kkjpsi$ events with invariant mass
between 4.4 and 5.2~GeV/$c^2$. The data agree well with the MC
simulation (shown as open histograms), indicating the existence of
signals that are produced from $ISR$.

We estimate the significance of the events between threshold and
6.0~GeV/$c^2$ by calculating the probability that the estimated
number of background events in the normalized $\jpsi$ sidebands
($12.3\pm 2.0$) fluctuates to the number of observed events in the
$\jpsi$ signal region ($93$) or more. It is found that the above
probability is very small, corresponding to a statistical
significance for the signal much larger than $10\sigma$.

\begin{figure}[htbp]
\psfig{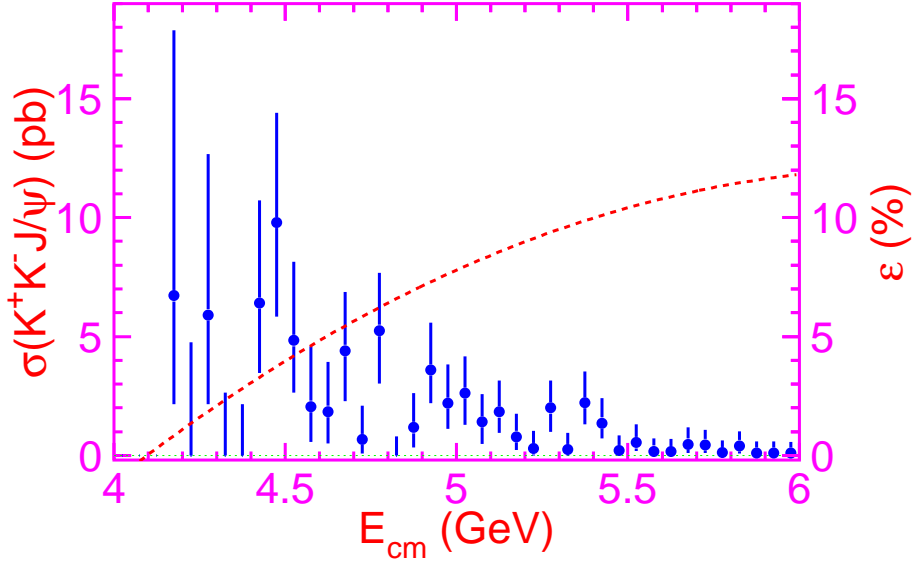} \caption{The
measured $\EE\to \kkjpsi$ cross section for CM energies between
threshold and 6.0~GeV (points with error bars). The errors are
statistical only; a 10\% systematic error that is common to all
the data points is not included. Bins without points have a
central value of zero. The dashed curve shows the energy-dependent
selection efficiency with scale in right-hand side.}
\label{xs_full}
\end{figure}

The $\EE\to \kkjpsi$ cross section for each $\kkjpsi$ mass bin
is computed with
 \beq
 \sigma_i = \frac{n^{\rm obs}_i - n_i^{\rm bkg}}
                 {\eff_i \lum_i \BR(\jpsi\to \LL)},
 \eeq
where $n^{\rm obs}_i$, $n_i^{\rm bkg}$, $\eff_i$, and $\lum_i$ are
the number of events observed in data, the number of background
events from a fit to the $\jpsi$ sideband events, the efficiency,
and the effective luminosity~\cite{kuraev} in the $i$-th $\kkjpsi$
mass bin, respectively~\cite{zerobin}. Due to the low statistics,
we fit the background distribution with a second-order polynomial
and take the fit number as the background in each $\kkjpsi$ mass
bin. The dilepton branching fraction, $\BR(\jpsi\to \LL)=11.87\%$,
is taken from Ref.~\cite{PDG}. The resulting cross sections are
shown in Fig.~\ref{xs_full}, where the error bars indicate the
combined statistical errors of the signal and the background
events~\cite{conrad}. The cross section we measure in the
4.25-4.30~GeV bin is consistent with the direct measurement at
$\sqrt{s}=4.26$~GeV by the CLEO experiment~\cite{cleoy4260}.

\begin{figure}[htbp]
\centerline{\psfig{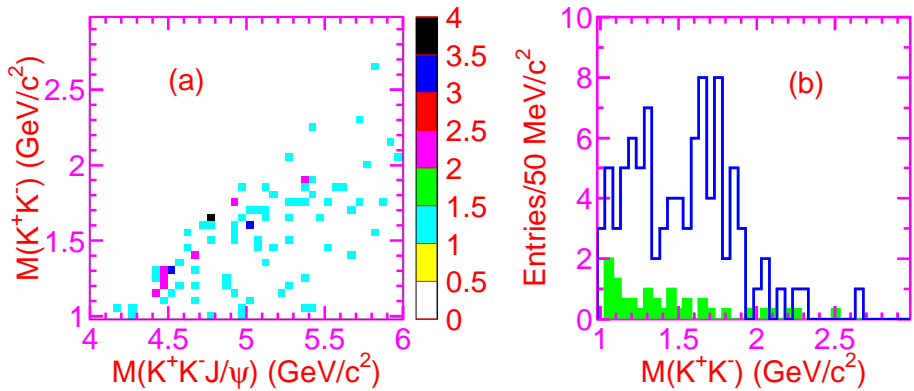}}
\caption{Scatter plot of $m_{\kk}$ versus $m_{\kkjpsi}$ (a) and
the projection on the $\kk$ invariant mass (b) for events in the
$\jpsi$ signal region. The shaded histogram is the distribution of
the normalized sideband events.} \label{scatter}
\end{figure}

There are a few sources of systematic errors for the cross section
measurement. The particle ID uncertainty, measured using the same
method as in Ref.~\cite{belley} with pure track samples, is 4\%;
the uncertainty in the tracking efficiency for tracks with angles
and momenta characteristic of signal events is about 1\%/track,
and is additive; efficiency uncertainties associated with the
$\jpsi$ mass and $\MMS$ requirements are also determined from a
study of the very pure $\EE\to \psp\to \ppjpsi$ event sample. In
this study we find that the detection efficiency is lower than
that inferred from the MC simulation by $(2.5\pm 0.4)\%$
relatively. A correction factor is applied to the final results
and 0.4\% is included in the systematic error. Belle measures the
luminosity with a precision of 1.4\%, and the uncertainty of the
$ISR$ photon radiator is 0.1\%~\cite{kuraev}. The main uncertainty
in the PHOKHARA generator~\cite{phokhara} is due to the modelling
of the $\kk$ mass spectrum. Figure~\ref{scatter} shows the $\kk$
invariant mass versus $\kkjpsi$ invariant mass, as well as the
projection on the $\kk$ invariant mass for events in the $\jpsi$
signal region. The $\kk$ invariant mass tends to be large and
close to the phase space boundary, with an accumulation of events
at 1.2~GeV/$c^2$ and 1.7~GeV/$c^2$. Simulations with modified
$\kk$ invariant mass distributions yield efficiencies that are
higher by 2-5\% for different $\kkjpsi$ masses. This is not
corrected for in the analysis, but is taken as the systematic
error (conservatively estimated as 5\%) for all $\kkjpsi$ mass
values. The angular distributions of the final state particles are
compared with the MC generation, no evidence was found for
non-$S$-wave components. Estimating the backgrounds using
different $\jpsi$ mass sidebands results in a change of background
events at the 0.18/50~MeV/$c^2$ level, corresponding to an average
of about a 6\% change in the cross section. According to the MC
simulation, the trigger efficiency for the final state is 98\%,
with an uncertainty that is smaller than 1\%. From
Ref.~\cite{PDG}, the uncertainty on the world averages for
$\BR(\jpsi\to \LL)=\BR(\jpsi\to \EE)+\BR(\jpsi\to \MM)$,
determined by linearly adding the errors for the $\EE$ and $\MM$
modes, is 1\%. Finally, the MC statistical error on the efficiency
is 1.5\%. These errors are summarized in Table~\ref{err_full}.
Assuming that all the sources are independent and adding them in
quadrature, we obtain a total systematic error on the cross
section of 10\%.

\begin{table}[htbp]
\caption{Systematic errors of the cross section measurement. They
are common for all data points.} \label{err_full}
\begin{tabular}{c  c}
\hline
  Source & Relative error (\%) \\\hline
 Particle ID &  4 \\
 Tracking efficiency & 4 \\
 $\jpsi$ mass and $\MMS$ selection & 0.4 \\
 Integrated luminosity & 1.4 \\
 $m_{\kk}$ distribution  & 5 \\
 Background estimation & 6 \\
 Trigger efficiency & 1 \\
 Branching fractions & 1 \\
 MC statistics & 1.5 \\
 \hline
 Sum in quadrature & 10 \\
 \hline
\end{tabular}
\end{table}

An unbinned maximum likelihood fit is applied to the $\kkjpsi$
mass spectrum in Fig.~\ref{mkkll_full}. Here the theoretical shape
is multiplied by the efficiency and effective luminosity, which
are functions of the $\kkjpsi$ invariant mass. The Breit-Wigner
function for a spin one resonance decaying into final state $f$
with mass ($M$), total width ($\Gamma_{\rm tot}$) and partial
width to $\EE$ ($\Gamma_{\EE}$) is
 \beq
\sigma(s)=\frac{M^2}{s}\frac{12\pi\Gamma_{\EE}\BR(R\to
f)\Gamma_{\rm tot}} {(s-M^2)^2+M^2\Gamma_{\rm
tot}^2}\frac{PS(\sqrt{s})}{PS(M)},
 \eeq
where $\BR(R\to f)$ is the branching fraction of the resonance to
final state $f$, and $PS(\sqrt{s})$ is the three-body decay phase
space factor for $X\to \kk\jpsi$. The MC-determined mass
resolution varies from 3~MeV/$c^2$ at $m_{\kkjpsi}= 4.3$~GeV/$c^2$
to 6.8~MeV/$c^2$ at 5.4~GeV/$c^2$. This is small compared to the
widths of the resonances in our study and is neglected.

We fit the $\kkjpsi$ invariant mass spectrum with one Breit-Wigner
plus a background term. The latter is a second-order polynomial
that is fit to the scaled sideband data. The dashed curve in
Fig.~\ref{mkkll_full} shows the fit results. The resonance
parameters are \( M = 4430^{+38}_{-43}~\hbox{MeV}/c^2 \), \(
\Gamma_{\rm tot} = 254^{+55}_{-46}~\hbox{MeV}/c^2 \), \(\BR(R\to
\kkjpsi)\cdot \Gamma_{\EE} = 1.9\pm 0.3~\hbox{eV}/c^2\), where the
errors are statistical only. Although the peak mass is close to
the $\psiftf$, $\Gamma_{\rm tot}$ is larger than its world average
of $62\pm 20$~MeV/$c^2$~\cite{PDG}. To determine the goodness of
fit, we bin the data so that the minimum expected number of events
in a bin is at least seven and determine a $\chi^2/ndf=10.7/6$,
corresponding to a confidence level (C.L.) of 10\%. Adding a
coherent $\y$ amplitude in the fit with mass and width fixed at
the Belle measurement~\cite{belley} yields an upper limit on
$\BR(\y\to \kkjpsi)\Gamma(\y\to \EE)<1.2$~eV/$c^2$ at 90\% C.L. A
fit using the above functions together with a coherent $\psiftf$
component with mass and width fixed at its world
average~\cite{PDG} values improves the fit~(solid curve in
Fig.~\ref{mkkll_full}) to $\chi^2/ndf=4.2/4$, C.L.=38\%; the
significance of the $\psiftf$ signal is found to be around
1.7$\sigma$ with a branching fraction for $\psiftf\to \kkjpsi$ at
the few per mille level; in this fit, the mass and width of the
second Breit-Wigner become \(4875\pm 132~\hbox{MeV}/c^2 \) and
\(630\pm 126~\hbox{MeV}/c^2 \), respectively.

\begin{figure}[htbp]
\psfig{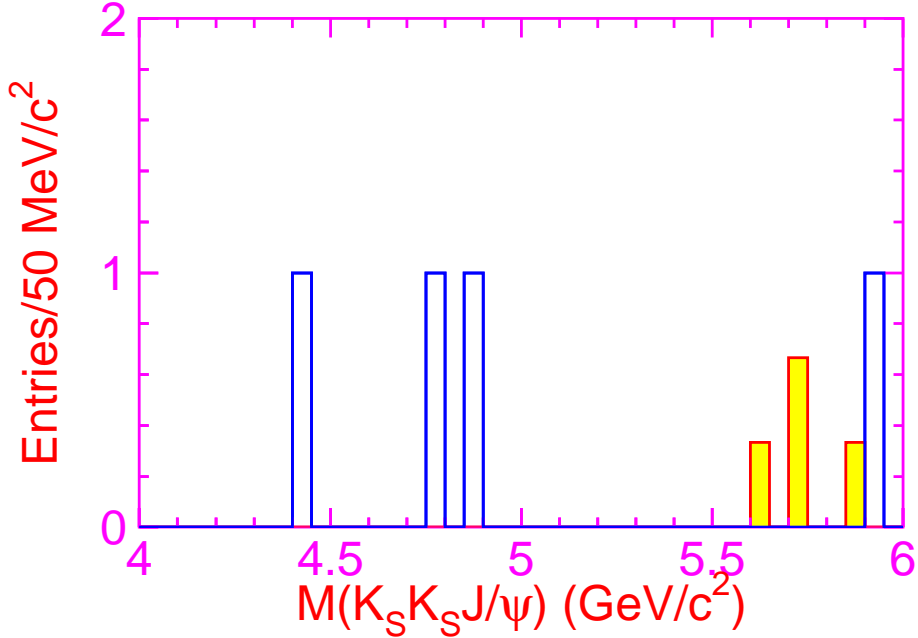}\caption{$\ksksjpsi$
invariant mass distribution of the selected $\EE\to \ksksjpsi$
candidates, blank histogram is for events in the $\jpsi$ signal
region, while the shaded histogram is the normalized sideband
backgrounds.} \label{ksksjpsi}
\end{figure}

We also search for $\EE\to \ksks\jpsi$ with the same data sample.
All the selection criteria are the same as for $\kkjpsi$ except
that the selection of $\kk$ is replaced by the selection of two
$\ks$'s decaying into $\pp$~\cite{ksrecon}. After selection, the
invariant mass distribution of the $\ksksjpsi$ candidates is shown
in Fig.~\ref{ksksjpsi}. Three events (one $\jpsi\to \EE$ and two
$\jpsi\to \MM$) are observed between 4.4 and 5.2~GeV/$c^2$ where a
large $\kkjpsi$ signal is observed. In the higher mass region, the
number of events in the $\jpsi$ signal region is about the same as
expected from the normalized sideband events.

MC simulation yields an average selection efficiency of
$\eff=(0.50\pm 0.04)\%$ for $m_{\ksksjpsi}\in
[4.4,5.2]~\hbox{GeV}/c^2$. Assuming that there is no background,
we obtain the average cross section for $\EE\to \ksksjpsi$ of \(
\overline{\sigma}_{\ksksjpsi} = 1.8^{+1.4}_{-1.1}~\hbox{pb} \),
where the error is statistical only. The average cross section for
$\EE\to \kkjpsi$ in the same mass range is \(
\overline{\sigma}_{\kkjpsi} = 3.1\pm 0.6~\hbox{pb} \), where the
error includes combined statistical and systematic uncertainties.
From the above two cross sections, we obtain \(
R=\frac{\overline{\sigma}_{\ksksjpsi}}{\overline{\sigma}_{\kkjpsi}}
= 0.6^{+0.5}_{-0.4} \), in agreement with the expectation
($R=1/2$) from isospin symmetry.

In summary, the process $\EE\to\kkjpsi$ is observed and the cross
section is measured for the CM energy between threshold and
6.0~GeV. There is one very broad structure; fits using either a
single Breit-Wigner function, or a $\psi(4415)$ plus a second
Breit-Wigner function yield resonance parameters that are very
different from those of the excited $\psi$ states currently listed
in Refs.~\cite{PDG,besres}. We observe two events near the $\y$
mass, with a cross section consistent with the CLEO
measurement~\cite{cleoy4260} at $\sqrt{s}=4.26$~GeV within the
large errors. We set an upper limit on $\BR(\y\to
\kkjpsi)\Gamma(\y\to \EE)<1.2$~eV/$c^2$ at 90\% C.L.

We thank the KEKB group for excellent operation of the
accelerator, the KEK cryogenics group for efficient solenoid
operations, and the KEK computer group and the NII for valuable
computing and Super-SINET network support. We acknowledge support
from MEXT and JSPS (Japan); ARC and DEST (Australia); NSFC, KIP of
CAS, and the 100 Talents program of CAS (China); DST (India);
MOEHRD, KOSEF and KRF (Korea); KBN (Poland); MES and RFAAE
(Russia); ARRS (Slovenia); SNSF (Switzerland); NSC and MOE
(Taiwan); and DOE (USA).


\begin{thebibliography}{**}

\bibitem{babary} BaBar Collaboration, B.~Aubert {\em et al.},
\Journal\PRL{95}{142001}{2005}.

\bibitem{cleoy} CLEO Collaboration, Q.~He {\em et al.},
\Journal\PRD{74}{091104(R)}{2006}.

\bibitem{belley} Belle Collaboration, C.~Z.~Yuan {\em et al.},
\Journal\PRL{99}{182004}{2007}.

\bibitem{babar_pppsp} BaBar Collaboration, B.~Aubert {\em et al.},
\Journal\PRL{98}{212001}{2007}.

\bibitem{belle_pppsp} Belle Collaboration, X.~L.~Wang {\em et al.},
\Journal\PRL{99}{142002}{2007}.

\bibitem{cleoy4260} CLEO Collaboration, T.E.~Coan {\em et al.},
\Journal\PRL{96}{162003}{2006}. 

\bibitem{Belle} Belle Collaboration, A.~Abashian {\em et al.},
Nucl. Instr. and Methods Phys. Res. Sect. A {\bf 479}, 117 (2002).

\bibitem{KEKB} S.~Kurokawa and E.~Kikutani,
Nucl. Instr. and Methods Phys. Res. Sect. A {\bf 499}, 1 (2003),
and other papers included in this volume.

\bibitem{phokhara} G.~Rodrigo, H.~Czy$\dot{\hbox{z}}$,
J.~H.~K$\ddot{\hbox{u}}$hn and M.~Szopa, Eur. Phys. J. C {\bf 24},
71 (2002).

\bibitem{besdist}BES Collaboration, J.~Z.~Bai {\em et al.},
\Journal\PRD{62}{032002}{2000}.

\bibitem{pid} E.~Nakano,
Nucl. Instr. and Methods Phys. Res. Sect. A {\bf 494}, 402 (2002).

\bibitem{footnote}
In this Letter, $m_{\kk\LL}-m_{\LL}+m_{\jpsi}$ is used instead of
the invariant mass of the four final state particles to improve
the mass resolution. Here $m_{\jpsi}$ is the nominal mass of
$\jpsi$.

\bibitem{kuraev} E.~A.~Kuraev and V.~S.~Fadin,
Sov.\ J.\ Nucl.\ Phys.\  {\bf 41}, 466 (1985) [Yad.\ Fiz.\ {\bf
41}, 733 (1985)].

\bibitem{zerobin} For the bins with no observed events, the number
of background events is taken to be zero when calculating the
68.3\% C.L. intervals.

\bibitem{PDG} Particle Data Group, W.-M.~Yao {\em et al.},
J. Phys. G {\bf 33}, 1 (2006).

\bibitem{conrad}J.~Conrad {\em et al.},
\Journal\PRD{67}{012002}{2003}.

\bibitem{ksrecon} F.~Fang, Ph.D thesis, University of Hawaii,
2003.

\bibitem{besres}BES Collaboration, M.~Ablikim {\em et al.},
  arXiv:0705.4500 [hep-ex].

\end{thebibliography}
\end{document}